\documentclass[fleqn,usenatbib]{mnras}
\pdfoutput=1
\usepackage{mathptmx}
\usepackage{txfonts}
\usepackage[T1]{fontenc}
\usepackage{savesym}
\savesymbol{iint}
\savesymbol{iiint}
\savesymbol{iiiint}
\savesymbol{idotsint}
\usepackage{amsmath}
\usepackage{graphicx}

\defcitealias{2015ApJ...809...29T}{TK}
\newcommand{\TK}{\citetalias{2015ApJ...809...29T}\ }
\defcitealias{Huangetal23}{ZH}
\newcommand{\ZH}{\citetalias{Huangetal23}\ }

\begin{document}

\title[Ultrarelativistic shocks]
{Particle acceleration at ultrarelativistic, perpendicular shock fronts}

\author[John G. Kirk, Brian Reville, Zhi-Qiu Huang]{John G. Kirk, Brian Reville, Zhi-Qiu Huang
\\
Max-Planck-Institut f\"ur Kernphysik, Postfach 10~39~80,
69029 Heidelberg, Germany}
\maketitle
\begin{abstract}
  Using an eigenfunction expansion to solve the transport equation,
  complemented by Monte-Carlo simulations,
  we show that ultrarelativistic shocks can be effective particle
  accelerators even when they fail to produce large amplitude
  turbulence in the downstream plasma. This finding contradicts the
  widely held belief that a uniform downstream magnetic field
  perpendicular to the shock normal inhibits acceleration by the first order Fermi
  process. In the ultrarelativistic limit, we find a
  stationary power-law particle spectrum of index \(s=4.17\) for these shocks, close to that
  predicted for a strictly parallel shock.
\end{abstract}

\begin{keywords}
  {radiation mechanisms: non-thermal --- acceleration of particles --- gamma rays: bursts}
\end{keywords}

\section{Introduction}
Discovered in the late 1970s, the theory of diffusive shock acceleration has
established itself as the primary mechanism discussed in
connection with the acceleration of cosmic rays,
and has also found many other applications.
The generalisation of this
mechanism to mildly relativistic shock fronts followed roughly a decade later,
and, in the early 2000's, it was found that
particles repeatedly crossing ultrarelativistic shocks can be accelerated
into a power law spectrum of index
\(s=4.23\)
\citep[for recent reviews, see][]{2014BrJPh..44..415B,2015SSRv..191..519S}.

However, in contrast to the nonrelativistic case, an ultrarelativistic
shock that overruns a region containing a uniform magnetic field is generically
superluminal, in the sense that its speed, when projected onto a magnetic field line, exceeds that of light.
At first sight, this poses a problem, since
acceleration requires repeated crossings of the shock front, and a particle downstream
of the shock cannot catch up with it by simply diffusing along a magnetic field line.
Since cross-field diffusion is generally 
strongly suppressed, this suggests 
that the relativistic extension of the
diffusive shock acceleration mechanism might be ineffective
\citep{1990ApJ...353...66B}.

This problem does not arise if the downstream field is effectively
scrambled by strong fluctuations on the scale of a gyroradius
\citep{2001MNRAS.328..393A}. However, a mechanism for producing such
fluctuations has so far not been identified.  For the case of shock
propagation into a weakly magnetized plasma,
where the Weibel instability is
thought responsible for the formation of the shock front,
particle-in-cell simulations have shown that acceleration is
facilitated by non-resonant scattering on the Weibel-induced
filaments \cite[e.g.][]{2009ApJ...698.1523S,2018MNRAS.477.5238P,2020Galax...8...33V}.  However, analytical considerations suggest
that scattering exclusively mediated by such
non-resonant interactions is not sufficient to provide the required
cross-field transport in the downstream plasma above a critical
particle energy
\citep{2010MNRAS.402..321L,2014MNRAS.439.2050R,2022ApJ...925..182H}.
Furthermore, such short length-scale fluctuations are susceptible to
damping in the hot downstream plasma, which further reduces the
critical energy \citep{2008ApJ...674..378C,2009ApJ...693L.127K,2013ApJ...771...54S,2015JPlPh..81a4501L}

In this paper, we re-assess these arguments by
solving a simple model of a relativistic, perpendicular
shock, thereby demonstrating quantitatively that particles are accelerated
into a power-law distribution whose index lies very close to that
predicted for the idealized, parallel shock case
\citep{2000ApJ...542..235K}.
In our model, energetic particles are assumed to 
diffuse in angle, whilst being deflected by a uniform
magnetic field that is {\em perpendicular} to the shock normal.
We solve this model for the case in which transport upstream
is scattering dominated, i.~e., the particles there are {\em unmagnetized},
whereas particles are {\em magnetized} when downstream and
follow essentially unscattered trajectories, gyrating about 
the regular, uniform magnetic field.
Two complementary techniques are employed.
First, we use a generalisation of the eigenfunction
approach \citep{2000ApJ...542..235K} that
takes into account the full, two-dimensional
anisotropy imposed on the particle distribution, including the
drift induced by the downstream magnetic
field. This technique is applied in the ultrarelativistic limit of large
shock Lorentz factor, \(\Gamma_{\rm s}\rightarrow\infty\),
which enables the eigenfunctions to be found in
closed form. The second method of solution is a Monte-Carlo simulation of many
individual trajectories, using a code described more fully in a
companion paper
\citep[][henceforth \lq\lq \citetalias{Huangetal23}\rq\rq]{Huangetal23}.
This is applied here to shocks ranging from mildly to highly relativistic
and
shows that, as \(\Gamma_{\rm s}\) rises from
\(2\) to \(50\), 
the power-law index
approaches the asymptotic value of
\(s=4.17\) found by the eigenfunction method. 
This value is close to, but slightly harder than the result obtained
previously for the case where scattering dominates both up and
downstream.

In section \ref{method} we formulate the equations describing particle
transport, and present details of
the first method of solution. Results of both methods, consisting of
the power law index of accelerated particles and the angular
dependence of their distribution function at the shock front, are
presented and compared
in section \ref{results}. In section \ref{discussion} we discuss
the physics underlying our assumptions and their range of applicability
and speculate on the implications of
the results for more realistic cases. 

\section{Method}
\label{method}
\subsection{Transport equation}
\label{methodupstream}
In the presence of isotropic
scattering in angle and gyration about a uniform field, the
transport equation governing the phase-space density \(f\)
of
relativistic particles is given
by eq~(1) of \cite{2015ApJ...809...29T}, (henceforth \lq\lq \citetalias{2015ApJ...809...29T}\rq\rq). Mixed coordinates are used in this
equation, with Cartesian coordinates in configuration
space measured in a frame in which the shock is
at rest in the plane \(x=0\). The upstream plasma occupies the region
\(x>0\), the downstream \(x<0\). Momentum space coordinates, on the other hand, are
expressed in spherical polar coordinates, measured in the
local (upstream or downstream) rest frame of the plasma.
In this paper, we depart from the notation of \TK
and use the shock normal as the axis for these
coordinates. 
Then, the momentum of an ultrarelativistic particle, in units of
\(c\,\times\,\)the rest mass, is
\(\vec{p}=\left(\gamma,\mu,\phi\right)\), with
\(\gamma\) the Lorentz factor, \(\arccos\mu\) the polar angle to the shock normal and \(\phi\)
the azimuthal angle about this axis. The
direction of motion of the shock front in this reference frame
is \(\mu=1\), 
and  \(\mu=0\), \(\phi=0\), 
is the direction of the uniform magnetic field
\(\vec{B}\).
Solutions are sought that are
stationary as seen in the shock rest frame, and
have no dependence on the
spatial coordinates \(y\) and \(z\).
eq~(12) of \TK is then
\begin{multline}
  \frac{2\Gamma^2\eta c}{\omega_{\rm g}}
  \left(\mu-\beta\right)\frac{\partial f}{\partial x}
  =\frac{\partial}{\partial\mu}\left(1-\mu^2\right)\frac{\partial f}{\partial\mu}
+ \frac{1}{1-\mu^2}\frac{\partial^2 f}{\partial\phi^2}\\
+\frac{2\Gamma\eta\mu\cos\phi}{\sqrt{1-\mu^2}}\frac{\partial f}{\partial\phi}
-2\Gamma\eta\sqrt{1-\mu^2}\sin\phi\frac{\partial f}{\partial\mu}\,,
\label{tk12}
\end{multline}
where \(c\beta\) is the shock velocity measured in the local rest frame of the plasma, 
\(\Gamma=\left(1-\beta^2\right)^{-1/2}\)
is the Lorentz factor of the
shock front and we have assumed \(\gamma\gg\Gamma\). 
The quantity \(\eta\) in this equation is the ratio of the gyro
frequency of an accelerated particle
in the uniform field, \(\omega_{\rm g}=\left|eB/\gamma mc\right|\),
to \(\Gamma\) times the scattering frequency. Following
\citet{2014MNRAS.439.2050R}, it can be written in terms
of the magnetization parameter \(\sigma_{\rm turb}\) associated with the strength \(\delta B\) of the magnetic fluctuations responsible for scattering,
their characteristic length scale \(\lambda\),
the magnetization parameter \(\sigma_{\rm reg}\) of the uniform, or {\em regular} upstream field and the
ion plasma frequency \(\omega_{\rm i}\):
\begin{align}
  \eta&=\left(\frac{m c}{m_{\rm p}\left<\gamma\right>\omega_{\rm i}\lambda}\right)\left(\frac{\sigma_{\rm reg}^{1/2}}{\sigma_{\rm turb}}\right)\left(\frac{\gamma}{\Gamma}\right)\,,
        \label{etaestimate}
\end{align} 
where \(m\) and \(m_{\rm p}\) are the rest masses of the accelerating particles and
that of the species dominating the plasma inertia respectively, and
\(\left<\gamma\right>m_{\rm p}c^2\) is the
mean energy per plasma particle.
When \(\eta\ll1\), scattering dominates the transport process and deflections by the regular magnetic field are unimportant.
We then refer to the particles as being unmagnetized. On the other hand, when \(\eta\gg1\), the particles are magnetized and
follow essentially unscattered trajectories in the uniform, regular magnetic field.
Equation~(\ref{tk12}) applies in both the upstream and the downstream regions, where
the speeds of the relativistic shock are denoted by \(\beta_{\rm s}\) and
\(\beta_{\rm d}\), respectively, and the corresponding
Lorentz factors are \(\Gamma_{\rm s}\gg1\) and \(\Gamma_{\rm d}\sim 1\).
Analogous notation, \(\eta_{\rm s,d}\), is used for the magnetization parameter.

 \subsection{Transport upstream}
As is the case for particles accelerated at a parallel, ultrarelativistic shock
front, we expect the stationary
particle distribution in the upstream medium to be
concentrated in a narrow cone of opening angle \(\sim 1/\Gamma_{\rm s}\) 
about the shock normal.
Consequently, it is convenient to
replace
\(\mu\) by the {\em stretched} variable
\begin{align}
\xi&=\left(1-\mu\right)/\left(1-\beta_{\rm s}\right)
\label{stretched}
\\
&\approx2\Gamma_{\rm s}^2\left(1-\mu\right)
\nonumber
\end{align}
\citep{1989A&A...225..559K}
which is zero for particles moving directly along the shock normal and unity
for those moving parallel to the shock front. 
The phase-space density is now to be regarded as a function of
\(x\) and the momentum space variables \((\gamma,\xi,\phi)\) that
lie in the domain \(\gamma>1\), \(0\le\xi<\infty\), \(0\le\phi<2\pi\)
which we denote by \(A\).
Then, inserting the definition (\ref{stretched}) into (\ref{tk12}),
expanding in powers of the small parameter \(1/\Gamma_{\rm s}\),
and assuming the particles are unmagnetized:
\begin{align}
  \eta_{\rm s}\ll 1\,,
  \label{etaupstream}
\end{align}
the transport equation upstream becomes
\begin{align}
\frac{\partial^2 f}{\partial \xi^2}+\frac{1}{\xi}\frac{\partial f}{\partial \xi}
+\frac{1}{4\xi^2}\frac{\partial^2 f}{\partial\phi^2}
-\left(\frac{1}{\xi}-1\right)\frac{\partial f}{\partial\hat{x}}
&=0\,,
\label{upstreameq}
\end{align}
where we have introduced the dimensionless
coordinate \(\hat{x}=x\left(4\Gamma_{\rm s}^2\omega_{\rm g}\right)/
\left(\eta c\right)\).

Equation (\ref{upstreameq}) is separable, resulting in an exponential dependence of \(f\) on \(\hat{x}\) and
two eigenvalue problems, one for the
\(\xi\)-dependence and one for the \(\phi\)-dependence. Taken together with the boundary
conditions (i) \(f\rightarrow 0\) as \(x\rightarrow\infty\), (ii) \(f\) bounded
at \(\xi=0\) and \(\infty\), (iii) \(f\) periodic in \(\phi\) with
period \(2\pi\) and (iv) \(f\) invariant under a change of sign in the
component of momentum parallel to the downstream magnetic field:
\(f(\gamma,\xi,\phi,x)=f(\gamma,\xi,\pi-\phi,x)\), each eigenvalue problem
is self-adjoint. Therefore, the solution can be
expanded in the two-dimensional eigenfunctions \(Q_i\):
\begin{align}
f\left(\gamma,\xi,\phi,x\right)
&=F(\gamma)
\sum_{i=1}^\infty\textrm{e}^{\Lambda_{i}\hat{x}}a_iQ_{i}\left(\xi,\phi\right)\,,
\label{expansion}
\end{align}
where the \(a_i\) are constants and the eigenfunctions
are orthogonal over the weighting function \(\xi-1\):
\begin{align}
\iint_{\vec{p}\in A}\!\!\!\!
\textrm{d}\xi\textrm{d}\phi\,
Q_i\left(\xi,\phi\right)
\left(\xi-1\right)Q_j\left(\xi,\phi\right)
&=0,\quad i\ne j.
\end{align}
In Appendix \ref{appendixA} we give explicit expressions for the eigenvalues
\(\Lambda_i\) and the eigenfunctions \(Q_{i}\).

\begin{figure}
	\begin{center}
		\includegraphics[scale=0.35]{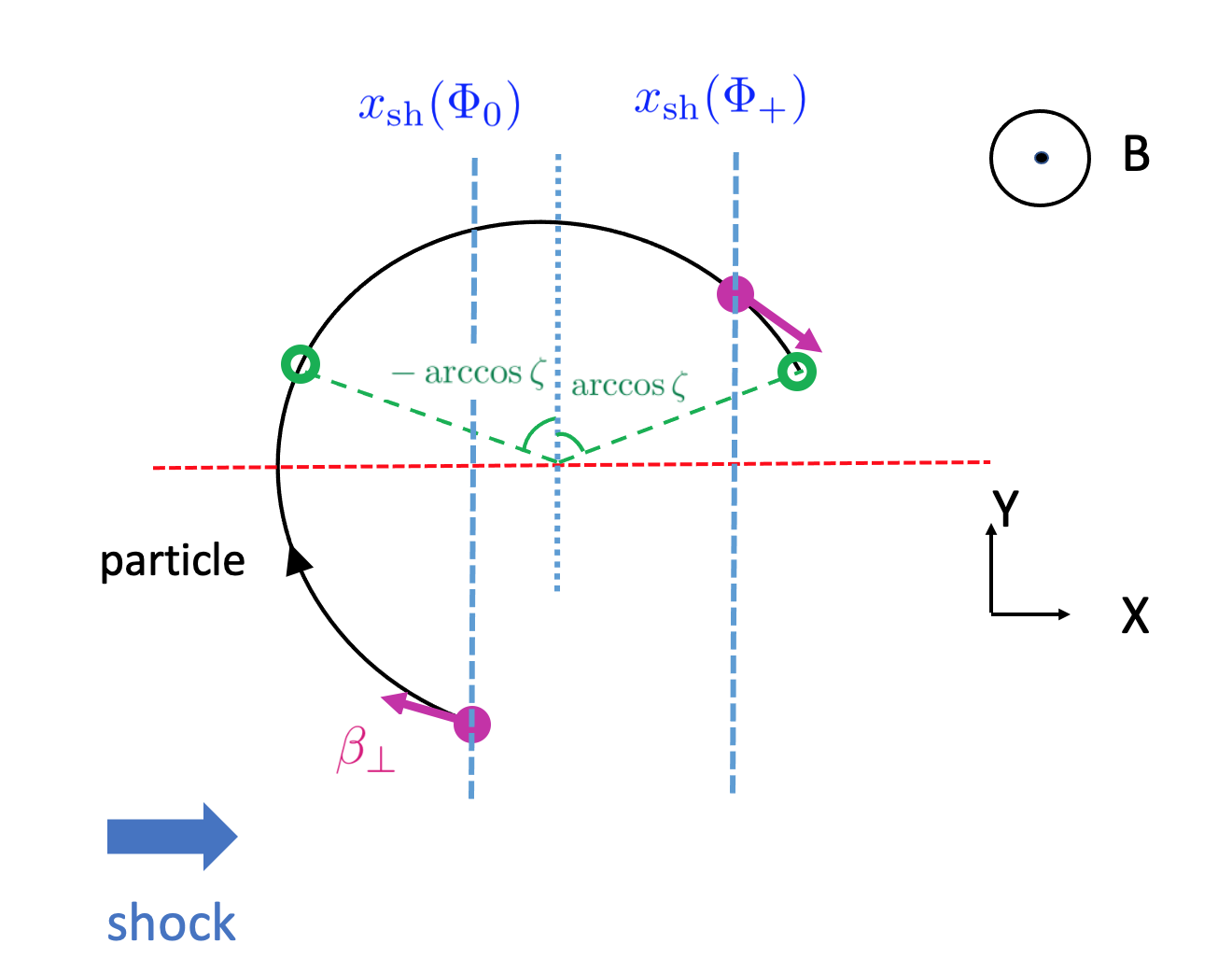}
		\caption{Sample trajectory for a positively charged particle with \(\vec{p}\in R\) as seen in the downstream frame. The shock velocity and magnetic field are directed along the positive \(x\) and \(z\) axes respectively. The magnitude of the particle's velocity \(\beta_{\perp}\) in the \(x-y\) plane is a constant of motion, and its direction along the positive \(x\) axis when the phase \(\Phi\) is an integer multiple of \(2\pi\). The filled magenta circles denote example positions where the particle enters and exits the downstream, at phases \(\Phi_0\) and \(\Phi_{+}\) respectively. To overtake the shock, the \(x\)-component of the
 particle velocity must exceed that of the shock, requiring \(-\textrm{arccos}\zeta<\Phi_{+}<\textrm{arccos}\zeta\), where \(\zeta = \beta_{\rm d}/\beta_\bot\). This range is delimited by the two green circles. For particles with \(\vec{p}\in U\), we make the replacements \(\Phi_0\rightarrow \Phi_{-}\) and \(\Phi_{+}\rightarrow \Phi_0\). Further details are provided in  Appendix \ref{appendixB}.
		\label{fig:schematic}}
	\end{center}
\end{figure}

      \begin{figure*}
        \includegraphics[width=0.9\textwidth]{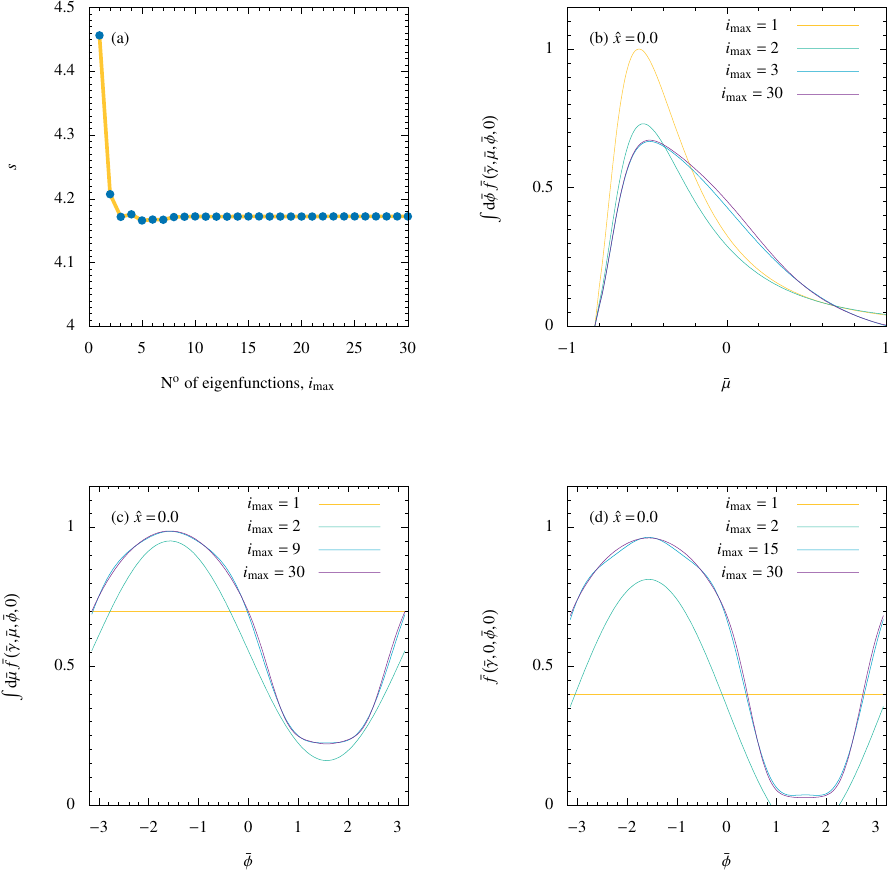}
        \caption{%
          \label{convergence}
          Convergence properties of the approximation scheme: (a) the
          power-law index \(s\) as a function of \(i_{\rm max}\),
          the number of eigenfunctions used in eq~(\ref{gimax}), (b) and (c)
          the phase space distribution \(\bar{f}\) at the shock front (\(\hat{x}=0\))
          as seen in the frame in which the shock front is at rest,  i.~e., at
          constant particle Lorentz factor in this frame (\(\bar{\gamma}\)).
          \(\bar{\mu}\) is the cosine of the angle between the
          particle momentum and the shock normal, \(\bar{\phi}\) the
          azimuthal angle about this axis. Particles entering the upstream
          have \(\bar{\mu}>0\).
          (b) shows \(\bar{f}\) averaged over phase,
          (c) \(\bar{f}\) averaged over \(\bar{\mu}\).
          (d) shows a slice of \( \bar{f} \) for particles
          that graze the shock, \(\bar{\mu}=0\). Results are plotted
          for \(i_{\rm max}=1,2\) and the fully converged \(i_{\rm max}=30\), as well as for an
          intermediate value that indicates the rapidity of convergence.
          In all cases the
          speed of the downstream plasma speed is
          \(\beta_{\rm d}=1/3\). In (b), (c) and (d) only the relative values of the
        functions plotted is physically significant.}
      \end{figure*}

\subsection{Transport downstream}
Transport in the downstream plasma is potentially more complex, since the 
accelerated particles are not concentrated in a narrow beam when
viewed from the frame in which the plasma is at rest. Here, we follow the
arguments presented by \cite{2014MNRAS.439.2050R}, who conclude that
deflection in the uniform magnetic field
perpendicular to the shock normal dominates the transport process for
particles with sufficiently large Lorentz factor, i.~e.,
\begin{align}
  \eta_{\rm d}\gg1\,.
  \label{etadownstream}
\end{align}
We concentrate
on these high energy particles, because those of lower
energy can be scattered in the turbulence
generated at the shock front, and there is general agreement
that the dominance of scattering on both
sides of a relativistic shock results in a power law spectrum
\(f\propto \gamma^{-s}\) with index \(s\approx4.2\).
      \begin{figure}
        \centering
        \includegraphics[width=0.45\textwidth]{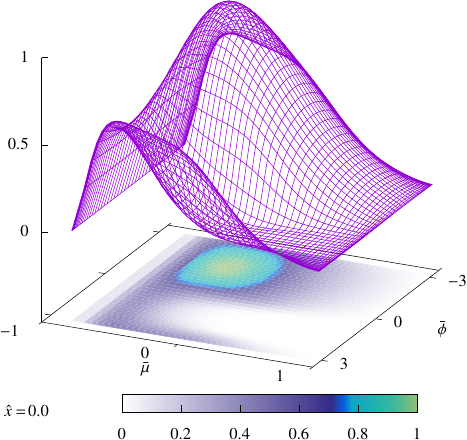}
        \caption{%
          The full angular distribution of the phase space density
          \(\bar{f}\left(\bar{\gamma},\bar{\mu},\bar{\phi}\right) \) of
          accelerated particles at the shock front (\(\hat{x}=0\)), as seen
          in the frame in which the shock front is at rest.
          The downstream magnetic field is in the direction \(\bar{\mu}=0\), \(\bar{\phi}=0\),
          particles crossing into the upstream region have \(\bar{\mu}>0\).
          \label{distribution}
}      \end{figure}

For \(\eta=\eta_{\rm d}\gg1\) and \(\Gamma=\Gamma_{\rm d}\sim1\), equation~(\ref{tk12}) reverts to Liouville's equation, albeit written in
our unconventional mixed coordinate system. The distribution of particles that are
too energetic to experience scattering downstream is, therefore, controlled
by Liouville's theorem, which dictates that
the phase-space density remains constant along
particle trajectories. When viewed from a frame of
reference in which the
downstream plasma is at rest 
these trajectories are simply helices along which the Lorentz factor \(\gamma\) remains
constant.

At the shock front, \(x=0\), the domain of momentum space \(A\) can be divided into
three non-overlapping sub-domains: (i) those trajectories crossing into
downstream that subsequently return to the shock front, denoted by \(R\)
(ii) those crossing into downstream that subsequently escape without
re-encountering the shock, denoted by \(E\), and (iii)
those crossing into upstream, (i.~e., all trajectories with \(0<\xi<1\))
denoted by \(U\). Given the coordinates in momentum space
\(\vec{p}=(\gamma,\xi,\phi)\) of a trajectory in \(R\), 
it is straightforward to compute the mapping \({\cal M}: R \rightarrow U\)
that relates the point \(\vec{p}\) on a particle trajectory to the
point \(\vec{p}_+=\left(\gamma_+,\xi_+,\phi_+\right)={\cal M}\vec{p}\) 
at which this trajectory returns to the shock front.
In Figure \ref{fig:schematic}, we provide a schematic sketch of
a returning trajectory
(for details see Appendix \ref{appendixB}).
Then, Liouville's theorem requires the distribution at the shock front, \(x=0\)
to satisfy
\begin{align}
f\left(\vec{p},0\right)&=f\left({\cal M}\vec{p},0\right)
\qquad \vec{p}\in R
.
\label{subsequent}
\end{align}
Conversely, given the coordinates of a trajectory that crosses
the shock from downstream to upstream,
one can simply invert this mapping to find the
coordinates
\(\vec{p}_-=\left(\gamma_-,\xi_-,\phi_-\right)={\cal M}^{-1}\vec{p}\) 
with which it previously entered the downstream region,
which implies
\begin{align}
f\left(\vec{p},0\right)&=f\left({\cal M}^{-1}\vec{p},0\right)
\qquad \vec{p}\in U \,.
\label{previous}
\end{align}
Liouville's theorem does not provide a constraint on
the points \(\vec{p}\in E\) on trajectories 
that escape from the shock into the downstream
plasma, but it is convenient to define the mapping \({\cal M}\) to be unity
when operating on these: \({\cal M}\vec{p}=\vec{p}\),
\(\vec{p}\in E\). 

\subsection{Approximation scheme}
\label{approximationscheme}
Since the problem, as formulated here,
does not contain a characteristic scale for
the particle Lorentz factor \(\gamma\), we look for solutions that are a power
law in this quantity,
\(F\left(\gamma\right)=\gamma^{-s}\).
The index \(s\) is then
determined by matching the phase-space density across the shock front, i.~e.,
by imposing the conditions (\ref{subsequent}) and (\ref{previous}) on the
expression (\ref{expansion})
for \(f\) in the upstream region, evaluated at the shock front \(\hat{x}=0\).

However, to find a numerical value for \(s\), it is necessary to implement
an approximation scheme. Here, we use a variant of the {\em Galerkin} method,
similar to that used by \cite{1987ApJ...315..425K}, which essentially truncates the
expansion (\ref{expansion}) after the first \(i_{\rm max}\) terms. Writing
\begin{align}
f\left(\vec{p},0\right)
&=\gamma^{-s}\left[g_{i_{\rm max}}\left(\xi,\phi\right)
                          + {\cal R}_{i_{\rm max}}\left(\xi,\phi\right)\right]\,,
  \label{galerkin1}\\
\noalign{\hbox{where}}
g_{i_{\rm max}}\left(\xi,\phi\right)&=\sum_{i=1}^{i_{\rm max}}
a_iQ_{i}\left(\xi,\phi\right) \,
\label{gimax}
\end{align}
is the desired approximation to the angular part of the distribution function at the shock front,
we demand that the residual \({\cal R}_{i_{\rm max}}\) be
orthogonal to the first \(i_{\rm max}\) eigenfunctions:
\begin{align}
\iint_{\vec{p}\in A}\textrm{d} \xi\textrm{d}\phi Q_{i}\left(\xi-1\right)
          {\cal R}_{i_{\rm max}}\left(\xi,\phi\right)
&=0\ \textrm{ for }i=1,\dots i_{\rm max}
\label{galerkin}
\end{align}
and, additionally, that the constraints (\ref{subsequent}) and (\ref{previous})
are satisfied not only by \(f\), but also by its approximation, \(\gamma^{-s}g_{i_{\rm max}}\).
This scheme can be motivated physically if the terms in the summation
in eq~(\ref{expansion}) are ordered by the eigenvalue \(\Lambda_i\),
such that
\begin{align}
  \Lambda_{i}&\ge\Lambda_{i+1},\qquad i\ge 1 .
  \label{ordering}
\end{align}
Since
\(\Lambda_1<0\), 
the higher order terms in the expansion decay ever more
rapidly with increasing distance from the shock in the upstream region.  The
particles described by these terms move almost in the plane of the
shock, and, as a consequence, receive only a small boost in energy in
a cycle of crossing and re-crossing.  It is, therefore, to be expected
that the index \(s\) is determined primarily by the first few terms in
the summation.

Multiplying equation~(\ref{galerkin1}) by
\(\gamma^s Q_j\) and the weighting function, and integrating over \(\vec{p}\in A\) leads to
\begin{align}
 \sum_{j=1}^{i_{\rm max}} \left[D_{ij}-M_{ij}(s)\right]a_j&=0\qquad i=1,\dots i_{\rm max}\,,
             \label{homogeneouseqs}
\end{align}
where the diagonal matrix \(D_{ij}\) results from substituting the
expansion (\ref{expansion}) into the left-hand side of (\ref{galerkin1}):
\begin{align}
    D_{ij}&=\iint_{\vec{p}\in A}  \!\! \textrm{d}\xi\textrm{d}\phi\,
Q_i\left(\xi,\phi\right)\left(\xi-1\right)Q_j\left(\xi,\phi\right)\,.
\end{align}
Similarly, the matrix \(M_{ij}(s)\) is obtained from the right-hand side by applying the constraints
(\ref{subsequent}) and (\ref{previous}) to the function \(g_{i_{\rm max}}\) and
the constraint
(\ref{galerkin}) to the function \( {\cal R}_{i_{\rm max}}\):
\begin{align}
  M_{ij}(s)&=\gamma^{s}
\iint_{\vec{p}\notin U}  \!\! \textrm{d}\xi\textrm{d}\phi\,
\gamma_+^{-s}Q_i\left(\xi,\phi\right)
\left(\xi-1\right)
Q_j\left(\xi_+,\phi_+\right)
\nonumber\\
&+
\gamma^{s}
\iint_{\vec{p}\in U} \!\! \textrm{d}\xi\textrm{d}\phi\,
\gamma_-^{-s}Q_i\left(\xi,\phi\right)
\left(\xi-1\right)
Q_j\left(\xi_-,\phi_-\right)\,.
\end{align}
Thus, eq~(\ref{homogeneouseqs}) represents a system of homogeneous, linear algebraic equations
for the \(a_j\), which have a non-trivial solution only if 
\begin{align}
\textrm{Det}\left[D_{ij}-M_{ij}(s)\right]&=0\,.
\label{determinant}
\end{align}
Given a guess for \(s\), it is straightforward to evaluate these matrices
by numerical quadrature. Then, a root-finder algorithm applied
to equation~(\ref{determinant}) yields \(s\) and the
coefficients \(a_i\) are found from the null-space of the
corresponding matrix.
  
      \section{Results}
\label{results}
The usefulness of the approximation scheme described in
\ref{approximationscheme} is confirmed by the rapid convergence as the
number of eigenfunctions \(i_{\rm max}\) is increased, as shown in
fig~\ref{convergence}.  The converged value of \(s\) for an
ultrarelativistic shock front in an ideal fluid, for which
\(\beta_{\rm d}=1/3\), is \(4.17\), close to, but slightly harder than
the result found when particle transport is dominated by scattering
both upstream and downstream. This value depends only on
\(\beta_{\rm d}\), and is, in particular, independent of both the
strength of the upstream scattering and the strength of the downstream
magnetic field. As shown in panel (a) of Fig~\ref{convergence}, 
convergence to this result requires only a few (\(\sim 4\))
eigenfunctions. 
Because of the ordering of the eigenfunctions, equation~(\ref{ordering}),
convergence of the angular dependence of the phase space density is slowest
at the shock front itself. Panels (b), (c) and (d) show this dependence
as seen in the frame in which the shock is at rest, and the flow is directed
along the shock normal. In this reference frame, the
transformed spherical polar coordinates are denoted by \(\bar{\gamma},\bar{\mu},\bar{\phi}\)
where
\begin{align}
  \bar{\gamma}&=\gamma\Gamma_{\rm s}\left(1 - \beta_{\rm s}\mu\right)\\
  \bar{\mu}&=\left(\mu - \beta_{\rm s}\right)/\left(1 - \mu\beta_{\rm s}\right)\\
  \bar{\phi}&=\phi
\end{align}
and, since the phase-space density, denoted in this frame by \(\bar{f}\), is a Lorentz invariant quantity
\begin{align}
  \bar{f}\left(\bar{\gamma},\bar{\mu},\bar{\phi},x\right)&=f\left(\gamma,\mu,\phi,x\right)\,.
\end{align}
Note 
that although
the average over either \(\bar{\phi}\) or \(\bar{\mu}\) of
the distribution converges with only \(\sim3\) and \(\sim9\) eigenfunctions, respectively, the
phase distribution of those particles that move precisely along the shock front
converges much more slowly, needing \(\sim15\) eigenfunctions. This illustrates the fact that
grazing particles have essentially no impact on the power law index.

      \begin{figure*}
        \includegraphics[width=0.8\textwidth]{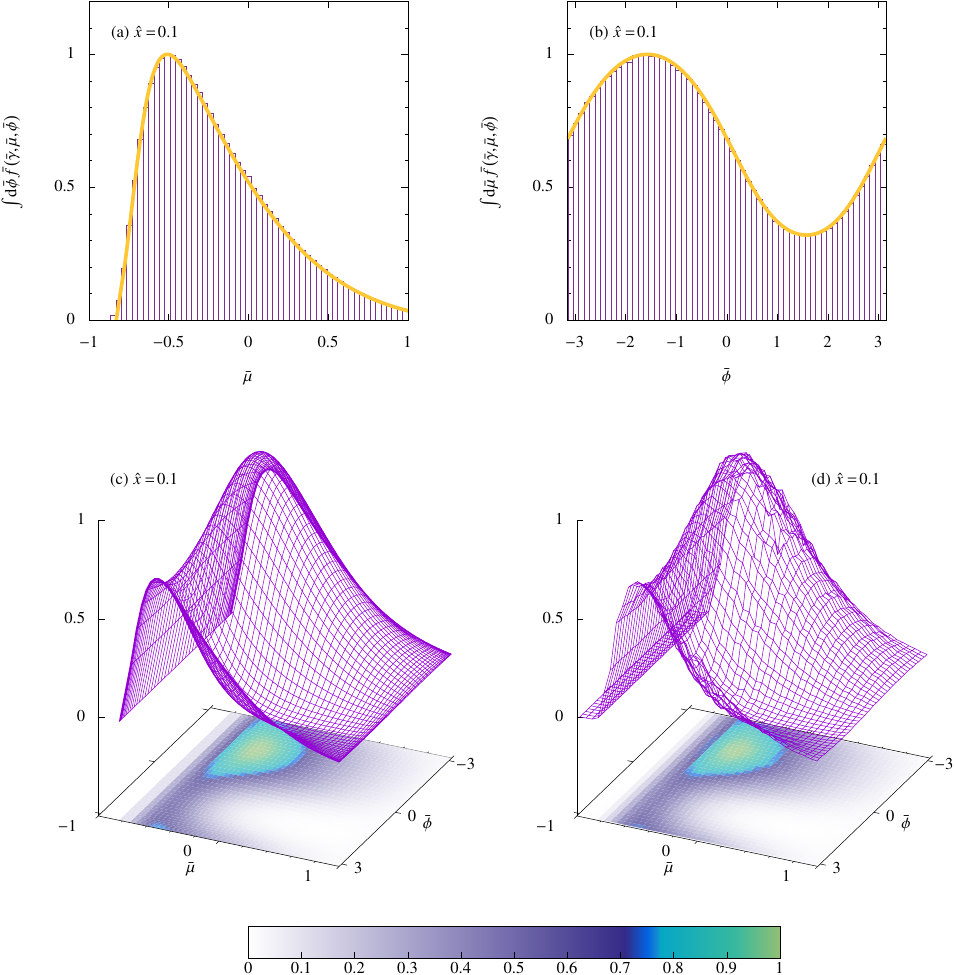}
        \caption{%
          \label{comparison}
          A comparison of the angular distributions just upstream of
          the shock front found by the eigenfunction method in the ultrarelativistic limit,
          and by Monte-Carlo simulation for \(\Gamma_{\rm s}=50\). In (a) the
          distributions averaged over phase (about the shock normal),
          and, in (b), over the angle to the
          shock normal are compared. In each case the result for 30
          eigenfunctions is shown as a yellow curve and that of the
          simulation as a purple histogram. In (c) and (d) the full
          distributions are shown for the eigenfunction and Monte Carlo
          methods, respectively.  }
      \end{figure*}

      \begin{figure}
        \includegraphics[width=0.45\textwidth]{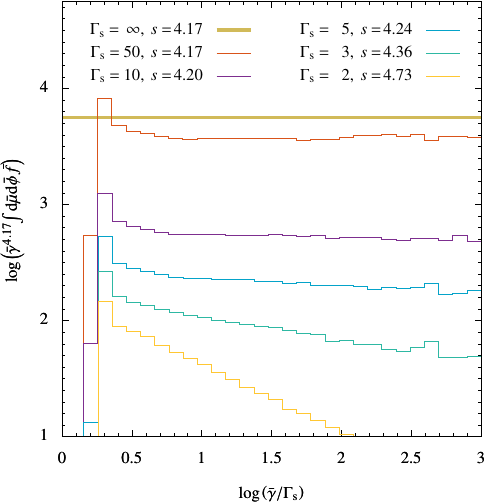}
        \caption{%
          \label{MCindex}
          The angle-averaged distribution just upstream at \(\hat{x}=0.1\),
          as a function of the particle Lorentz factor \(\bar{\gamma}\),
          measured in the rest frame of the shock. As the shock Lorentz factor
          \(\Gamma_{\rm s}\) increases,
          the power-law section of the distribution converges
          towards the result found by the eigenfunction method in the
          ultrarelativistic limit, shown here as a horizontal line.
          For each \(\Gamma_{\rm s}\), the power-law index \(s\), found from a least-squares fit to the data
          in the region \(0.5<\bar{\gamma}/\Gamma_{\rm s}<3\), is listed in the legend.}
      \end{figure}

The angular distribution at the shock front is, as expected,
anisotropic. In addition to the anisotropy with respect to the shock
normal, which is well-known from earlier studies of the
scattering-dominated case and which arises because the relative
velocity of the upstream and downstream plasmas is only slightly less
than the particle velocity, a strong anisotropy in the azimuthal angle \(\bar{\phi}\)
is present because the downstream magnetic field imposes a drift along
the shock front. This is clearly seen in the full 2-D angular
distribution at the shock front, as shown Fig.~\ref{distribution}. In terms of a right-handed
system of coordinates with the magnetic field along the positive
\(z\)-axis, the drift for a positively charged particle is in the
positive \(y\)-direction (\(\sin\bar{\phi}<0\)), i.~e., in the direction opposite to that
of the electric
field seen in the shock rest frame. Thus, as seen in the rest frame of the shock, particles gain
energy during an excursion upstream, but lose some on their downstream loop.

Monte-Carlo simulations, using the code described in \ZH, provide an
independent cross-check of these results, as well as extending them by
lifting the restriction to the ultrarelativistic limit.  In
Fig~\ref{comparison}, we compare the angular distributions found by
both methods. Shock grazing particles are problematic in the Monte-Carlo approach when sampling the distribution precisely on the shock
surface. Making the comparison a short distance upstream,
here at the surface \(\hat{x}=0.1\), mitigates the problem.
Results for the eigenfunction method are shown in the
ultrarelativistic limit, since using a finite value of \(\Gamma_{\rm s}\)
would necessitate a numerical evaluation of the
eigenfunctions. The Monte-Carlo simulations were performed
with \(\Gamma_{\rm s}=50\). Agreement is generally excellent, with
small fluctuations visible only in the Monte-Carlo results for the full 2-D distribution.
  
The distribution of accelerated particles in energy for shock speeds ranging from mildly to highly relativistic
is shown in Fig~\ref{MCindex}. These are found using Monte-Carlo simulations
that inject \(10^6\) particles into the upstream with 
\(\gamma=2\Gamma_{\rm s}^2\), and a uniform angular distribution within a cone
about the shock normal of opening angle \(1/\Gamma_{\rm s}\). The trajectories
are then followed until they escape downstream, whilst
registering the value of \(\bar{\gamma}\) at each crossing of the shock. After several crossings,
the distribution settles into a power law, that extends up to a point at which the statistical
noise becomes significant. 
For simplicity, we choose in each case the jump
conditions for a relativistic gas: \(\beta_{\rm d}\beta_{\rm s}=1/3\).
The figure shows
the distribution weighted by the factor \(\bar{\gamma}^{4.17}\)
in order to highlight the departure of the finite \(\Gamma_{\rm s}\) results
from the ultrarelativistic result found by the eigenfunction method. It
can be seen that the power-law index \(s\)
is within a few percent of its asymptotic value
for \(\Gamma_{\rm s}>5\).

\section{Discussion}
\label{discussion}

Whether or not the mechanism of diffusive shock acceleration operates
effectively at a perpendicular shock is a question that is still the
subject of controversy, over four decades after the publication of the
discovery papers, which implicitly addressed parallel shocks. In this
context \lq\lq effectively\rq\rq\ means either that the acceleration
rate is comparable to or faster than that at a parallel shock, or that
the power-law index of the stationary phase-space distribution of
accelerated particles is close to or harder than that produced at a
parallel shock. The main result of this paper concerns the power-law index
produced by highly relativistic, and, therefore, generically perpendicular
shocks. Assuming that particles can be
treated as unmagnetized when upstream of the shock
(\ref{etaupstream}), but as magnetized when downstream
(\ref{etadownstream}), we demonstrate quantitatively that these shocks
are just as effective accelerators as the
possibly less realistic parallel shocks addressed by previous
analytical work. 
This result has major implications also for the expected acceleration rate, and the
related maximum energy to which particles can be accelerated
at a shock in a given
physical situation, questions which are 
addressed in \ZH. 

The persistence of turbulence in the {\em downstream\/} medium at a level
sufficient to demagnetize particles is generally perceived to be a major
problem for the theory of diffusive shock acceleration at relativistic shocks
\citep[e.~g.,][]{2012SSRv..173..309B}. 
Our assumption that turbulence is completely negligible downstream is
specifically designed to address this point. It is, therefore, not restrictive, in the sense that
diffusive acceleration can be expected to proceed as previously predicted
if this assumption is not justified in a particular application. 

On the other hand, the assumption that particles diffuse in angle when {\em upstream\/} of the shock is
important. Unless a degree of randomness enters into the trajectories that return to
the shock from upstream, diffusive acceleration will cease
and particles will receive
only a single, finite boost in energy before being swept away
downstream \citep{1990ApJ...353...66B,2009MNRAS.393..587P}. We have adopted the simple prescription
of isotropic diffusion to describe this randomness, since it
renders the analytic approximations tractable; 
at least in the parallel shock case, this assumption does not appear to be crucial
\citep{2000ApJ...542..235K}. 
However, the validity of our approach does depend on the presence of turbulence of sufficient amplitude
in the upstream medium.
In the case of a weakly magnetized plasma \(\sigma_{\rm reg}< 10^{-3}\),
such as might be encountered by the blast wave of a gamma-ray burst (GRB),
a lower limit on the level of upstream turbulence is provided by shock-generated Weibel filaments,
which, according to particle-in-cell
simulations and analytical considerations, are amplified to \(\sigma_{\rm turb}\approx 0.1\).
According to (\ref{etaestimate}), this is already sufficient to demagnetize
particles with \(\gamma<0.1\times\Gamma_{\rm s}\left({m_{\rm p}/m}\right)\sigma_{\rm reg}^{-1/2}\),
and, therefore, enable diffusive acceleration even if the filaments are strongly damped downstream.
However, this limit is unduly restrictive if turbulence exists in the upstream that is not
directly excited by the processes that form the shock. In the case of a GRB, for example, 
the ionization, heating and pair-loading of the surrounding medium by the prompt emission
\citep{2002ApJ...565..808B,2022ApJ...933...74G}
seems unlikely to leave behind a quiescent environment.
Furthermore, turbulence is
known to be present in the winds of the progenitors
of some supernovae of the type~Ibc \citep{2012ApJ...752...17W} that is associated with
long duration GRBs.
On the other hand, if the upstream medium is strongly magnetized, such as 
in the case of the
termination shock of a pulsar wind, turbulence is embedded in the outflow by the
pulsar
and is thought to facilitate acceleration \citep{2011ApJ...741...39S,2017ApJ...835..235G}.
Whereas previous discussions assumed this process to be confined to the equatorial region of
the wind, where the regular field component vanishes, the results presented above
suggest that it may persist to higher latitudes, with interesting implications for modelling
the emission from pulsar wind nebulae
\citep{2015MNRAS.449.3149O}.

In summary, if sufficient turbulence is present in the upstream medium, 
our results demonstrate that the perpendicular magnetic field in the plasma
downstream of a relativistic shock front does not inhibit acceleration. Rather than being swept away
without returning to the shock front, particles indeed return,
and populate a power-law distribution whose index is
insensitive to the nature of the dominant transport mechanism
downstream.

\section*{Acknowledgements}
We thank Nils Schween, Gwenael Giacinti,
Tony Bell, Martin Lemoine and Daniel Gro{\v{s}}elj for helpful discussions.
\section*{Data availability}
No data relevant to this paper are available.

\bibliographystyle{mnras}

\begin{thebibliography}{}
	\makeatletter
	\relax
	\def\mn@urlcharsother{\let\do\@makeother \do\$\do\&\do\#\do\^\do\_\do\%\do\~}
	\def\mn@doi{\begingroup\mn@urlcharsother \@ifnextchar [ {\mn@doi@}
		{\mn@doi@[]}}
	\def\mn@doi@[#1]#2{\def\@tempa{#1}\ifx\@tempa\@empty \href
		{http://dx.doi.org/#2} {doi:#2}\else \href {http://dx.doi.org/#2} {#1}\fi
		\endgroup}
	\def\mn@eprint#1#2{\mn@eprint@#1:#2::\@nil}
	\def\mn@eprint@arXiv#1{\href {http://arxiv.org/abs/#1} {{\tt arXiv:#1}}}
	\def\mn@eprint@dblp#1{\href {http://dblp.uni-trier.de/rec/bibtex/#1.xml}
		{dblp:#1}}
	\def\mn@eprint@#1:#2:#3:#4\@nil{\def\@tempa {#1}\def\@tempb {#2}\def\@tempc
		{#3}\ifx \@tempc \@empty \let \@tempc \@tempb \let \@tempb \@tempa \fi \ifx
		\@tempb \@empty \def\@tempb {arXiv}\fi \@ifundefined
		{mn@eprint@\@tempb}{\@tempb:\@tempc}{\expandafter \expandafter \csname
			mn@eprint@\@tempb\endcsname \expandafter{\@tempc}}}
	
	\bibitem[\protect\citeauthoryear{{Achterberg}, {Gallant}, {Kirk}  \&
		{Guthmann}}{{Achterberg} et~al.}{2001}]{2001MNRAS.328..393A}
	{Achterberg} A.,  {Gallant} Y.~A.,  {Kirk} J.~G.,   {Guthmann} A.~W.,  2001,
	\mn@doi [\mnras] {10.1046/j.1365-8711.2001.04851.x}, \href
	{https://ui.adsabs.harvard.edu/abs/2001MNRAS.328..393A} {328, 393}
	
	\bibitem[\protect\citeauthoryear{{Begelman} \& {Kirk}}{{Begelman} \&
		{Kirk}}{1990}]{1990ApJ...353...66B}
	{Begelman} M.~C.,  {Kirk} J.~G.,  1990, \mn@doi [\apj] {10.1086/168590}, \href
	{http://adsabs.harvard.edu/abs/1990ApJ...353...66B} {353, 66}
	
	\bibitem[\protect\citeauthoryear{{Bell}}{{Bell}}{2014}]{2014BrJPh..44..415B}
	{Bell} A.~R.,  2014, \mn@doi [Brazilian Journal of Physics]
	{10.1007/s13538-014-0219-5}, \href
	{https://ui.adsabs.harvard.edu/abs/2014BrJPh..44..415B} {44, 415}
	
	\bibitem[\protect\citeauthoryear{{Beloborodov}}{{Beloborodov}}{2002}]{2002ApJ...565..808B}
	{Beloborodov} A.~M.,  2002, \mn@doi [\apj] {10.1086/324195}, \href
	{https://ui.adsabs.harvard.edu/abs/2002ApJ...565..808B} {565, 808}
	
	\bibitem[\protect\citeauthoryear{{Bykov}, {Gehrels}, {Krawczynski}, {Lemoine},
		{Pelletier}  \& {Pohl}}{{Bykov} et~al.}{2012}]{2012SSRv..173..309B}
	{Bykov} A.,  {Gehrels} N.,  {Krawczynski} H.,  {Lemoine} M.,  {Pelletier} G.,
	{Pohl} M.,  2012, \mn@doi [\ssr] {10.1007/s11214-012-9896-y}, \href
	{https://ui.adsabs.harvard.edu/abs/2012SSRv..173..309B} {173, 309}
	
	\bibitem[\protect\citeauthoryear{{Chang}, {Spitkovsky}  \& {Arons}}{{Chang}
		et~al.}{2008}]{2008ApJ...674..378C}
	{Chang} P.,  {Spitkovsky} A.,   {Arons} J.,  2008, \mn@doi [\apj]
	{10.1086/524764}, \href
	{https://ui.adsabs.harvard.edu/abs/2008ApJ...674..378C} {674, 378}
	
	\bibitem[\protect\citeauthoryear{{Giacch{\`e}} \& {Kirk}}{{Giacch{\`e}} \&
		{Kirk}}{2017}]{2017ApJ...835..235G}
	{Giacch{\`e}} S.,  {Kirk} J.~G.,  2017, \mn@doi [\apj]
	{10.3847/1538-4357/835/2/235}, \href
	{https://ui.adsabs.harvard.edu/abs/2017ApJ...835..235G} {835, 235}
	
	\bibitem[\protect\citeauthoryear{{Gro{\v{s}}elj}, {Sironi}  \&
		{Beloborodov}}{{Gro{\v{s}}elj} et~al.}{2022}]{2022ApJ...933...74G}
	{Gro{\v{s}}elj} D.,  {Sironi} L.,   {Beloborodov} A.~M.,  2022, \mn@doi [\apj]
	{10.3847/1538-4357/ac713e}, \href
	{https://ui.adsabs.harvard.edu/abs/2022ApJ...933...74G} {933, 74}
	
	\bibitem[\protect\citeauthoryear{{Huang}, {Kirk}, {Giacinti}  \&
	{Reville}}{{Huang} et~al.}{2022}]{2022ApJ...925..182H}
	{Huang} Z.-Q.,  {Kirk} J.~G.,  {Giacinti} G.,   {Reville} B.,  2022, \mn@doi
	[\apj] {10.3847/1538-4357/ac3f38}, \href
	{https://ui.adsabs.harvard.edu/abs/2022ApJ...925..182H} {925, 182}
	
	\bibitem[\protect\citeauthoryear{{Huang}, {Reville}, {Kirk}  \&
		{Giacinti}}{{Huang} et~al.}{2023}]{Huangetal23}
	{Huang} Z.-Q.,  {Reville} B.,  {Kirk} J.~G.,   {Giacinti} G.,  2023, to be
	submitted to \mnras
	
	\bibitem[\protect\citeauthoryear{{Ince}}{{Ince}}{1956}]{1927ode..book.....I}
	{Ince} E.~L.,  1956, {Ordinary Differential Equations}.
	Dover Publications, Inc, New York
	
	\bibitem[\protect\citeauthoryear{{Keshet}, {Katz}, {Spitkovsky}  \&
		{Waxman}}{{Keshet} et~al.}{2009}]{2009ApJ...693L.127K}
	{Keshet} U.,  {Katz} B.,  {Spitkovsky} A.,   {Waxman} E.,  2009, \mn@doi
	[\apjl] {10.1088/0004-637X/693/2/L127}, \href
	{https://ui.adsabs.harvard.edu/abs/2009ApJ...693L.127K} {693, L127}
	
	\bibitem[\protect\citeauthoryear{{Kirk} \& {Schneider}}{{Kirk} \&
		{Schneider}}{1987}]{1987ApJ...315..425K}
	{Kirk} J.~G.,  {Schneider} P.,  1987, \mn@doi [\apj] {10.1086/165147}, \href
	{https://ui.adsabs.harvard.edu/abs/1987ApJ...315..425K} {315, 425}
	
	\bibitem[\protect\citeauthoryear{{Kirk} \& {Schneider}}{{Kirk} \&
		{Schneider}}{1989}]{1989A&A...225..559K}
	{Kirk} J.~G.,  {Schneider} P.,  1989, \aap, \href
	{https://ui.adsabs.harvard.edu/abs/1989A&A...225..559K} {225, 559}
	
	\bibitem[\protect\citeauthoryear{{Kirk}, {Guthmann}, {Gallant}  \&
		{Achterberg}}{{Kirk} et~al.}{2000}]{2000ApJ...542..235K}
	{Kirk} J.~G.,  {Guthmann} A.~W.,  {Gallant} Y.~A.,   {Achterberg} A.,  2000,
	\mn@doi [\apj] {10.1086/309533}, \href
	{https://ui.adsabs.harvard.edu/abs/2000ApJ...542..235K} {542, 235}
	
	\bibitem[\protect\citeauthoryear{{Lemoine}}{{Lemoine}}{2015}]{2015JPlPh..81a4501L}
	{Lemoine} M.,  2015, \mn@doi [Journal of Plasma Physics]
	{10.1017/S0022377814000920}, \href
	{https://ui.adsabs.harvard.edu/abs/2015JPlPh..81a4501L} {81, 455810101}
	
	\bibitem[\protect\citeauthoryear{{Lemoine} \& {Pelletier}}{{Lemoine} \&
		{Pelletier}}{2010}]{2010MNRAS.402..321L}
	{Lemoine} M.,  {Pelletier} G.,  2010, \mn@doi [\mnras]
	{10.1111/j.1365-2966.2009.15869.x}, \href
	{https://ui.adsabs.harvard.edu/abs/2010MNRAS.402..321L} {402, 321}
	
	\bibitem[\protect\citeauthoryear{{Olmi}, {Del Zanna}, {Amato}  \&
		{Bucciantini}}{{Olmi} et~al.}{2015}]{2015MNRAS.449.3149O}
	{Olmi} B.,  {Del Zanna} L.,  {Amato} E.,   {Bucciantini} N.,  2015, \mn@doi
	[\mnras] {10.1093/mnras/stv498}, \href
	{https://ui.adsabs.harvard.edu/abs/2015MNRAS.449.3149O} {449, 3149}
	
	\bibitem[\protect\citeauthoryear{{Pelletier}, {Lemoine}  \&
		{Marcowith}}{{Pelletier} et~al.}{2009}]{2009MNRAS.393..587P}
	{Pelletier} G.,  {Lemoine} M.,   {Marcowith} A.,  2009, \mn@doi [\mnras]
	{10.1111/j.1365-2966.2008.14219.x}, \href
	{https://ui.adsabs.harvard.edu/abs/2009MNRAS.393..587P} {393, 587}
	
	\bibitem[\protect\citeauthoryear{{Plotnikov}, {Grassi}  \& {Grech}}{{Plotnikov}
		et~al.}{2018}]{2018MNRAS.477.5238P}
	{Plotnikov} I.,  {Grassi} A.,   {Grech} M.,  2018, \mn@doi [\mnras]
	{10.1093/mnras/sty979}, \href
	{https://ui.adsabs.harvard.edu/abs/2018MNRAS.477.5238P} {477, 5238}
	
	\bibitem[\protect\citeauthoryear{{Reville} \& {Bell}}{{Reville} \&
		{Bell}}{2014}]{2014MNRAS.439.2050R}
	{Reville} B.,  {Bell} A.~R.,  2014, \mn@doi [\mnras] {10.1093/mnras/stu088},
	\href {https://ui.adsabs.harvard.edu/abs/2014MNRAS.439.2050R} {439, 2050}
	
	\bibitem[\protect\citeauthoryear{{Sironi} \& {Spitkovsky}}{{Sironi} \&
		{Spitkovsky}}{2009}]{2009ApJ...698.1523S}
	{Sironi} L.,  {Spitkovsky} A.,  2009, \mn@doi [\apj]
	{10.1088/0004-637X/698/2/1523}, \href
	{http://adsabs.harvard.edu/abs/2009ApJ...698.1523S} {698, 1523}
	
	\bibitem[\protect\citeauthoryear{{Sironi} \& {Spitkovsky}}{{Sironi} \&
		{Spitkovsky}}{2011}]{2011ApJ...741...39S}
	{Sironi} L.,  {Spitkovsky} A.,  2011, \mn@doi [\apj]
	{10.1088/0004-637X/741/1/39}, \href
	{http://ads.nao.ac.jp/abs/2011ApJ...741...39S} {741, 39}
	
	\bibitem[\protect\citeauthoryear{{Sironi}, {Spitkovsky}  \& {Arons}}{{Sironi}
		et~al.}{2013}]{2013ApJ...771...54S}
	{Sironi} L.,  {Spitkovsky} A.,   {Arons} J.,  2013, \mn@doi [\apj]
	{10.1088/0004-637X/771/1/54}, \href
	{https://ui.adsabs.harvard.edu/abs/2013ApJ...771...54S} {771, 54}
	
	\bibitem[\protect\citeauthoryear{{Sironi}, {Keshet}  \& {Lemoine}}{{Sironi}
		et~al.}{2015}]{2015SSRv..191..519S}
	{Sironi} L.,  {Keshet} U.,   {Lemoine} M.,  2015, \mn@doi [\ssr]
	{10.1007/s11214-015-0181-8}, \href
	{https://ui.adsabs.harvard.edu/abs/2015SSRv..191..519S} {191, 519}
	
	\bibitem[\protect\citeauthoryear{{Takamoto} \& {Kirk}}{{Takamoto} \&
		{Kirk}}{2015}]{2015ApJ...809...29T}
	{Takamoto} M.,  {Kirk} J.~G.,  2015, \mn@doi [\apj]
	{10.1088/0004-637X/809/1/29}, \href
	{https://ui.adsabs.harvard.edu/abs/2015ApJ...809...29T} {809, 29}
	
	\bibitem[\protect\citeauthoryear{{Vanthieghem}, {Lemoine}, {Plotnikov},
		{Grassi}, {Grech}, {Gremillet}  \& {Pelletier}}{{Vanthieghem}
		et~al.}{2020}]{2020Galax...8...33V}
	{Vanthieghem} A.,  {Lemoine} M.,  {Plotnikov} I.,  {Grassi} A.,  {Grech} M.,
	{Gremillet} L.,   {Pelletier} G.,  2020, \mn@doi [Galaxies]
	{10.3390/galaxies8020033}, \href
	{https://ui.adsabs.harvard.edu/abs/2020Galax...8...33V} {8, 33}
	
	\bibitem[\protect\citeauthoryear{{Wellons}, {Soderberg}  \&
		{Chevalier}}{{Wellons} et~al.}{2012}]{2012ApJ...752...17W}
	{Wellons} S.,  {Soderberg} A.~M.,   {Chevalier} R.~A.,  2012, \mn@doi [\apj]
	{10.1088/0004-637X/752/1/17}, \href
	{https://ui.adsabs.harvard.edu/abs/2012ApJ...752...17W} {752, 17}
	
	\makeatother
\end{thebibliography}

\appendix
    \section{2D eigenfunctions}
\label{appendixA}
Inserting the expansion (\ref{expansion}) into (\ref{upstreameq})
and separating the variables according to
\begin{align}
Q_i\left(\xi,\phi\right)&=T_{i}\left(\xi\right)S_i\left(\phi\right)
\end{align}
one obtains for the \(\phi\)-dependent function
\begin{align}
S''_i&=-j^2S_i\,,
\end{align}
where \(j\) is a constant, which, since \(S\) is 
periodic with period \(2\pi\), is an integer. Without loss
of generality, we can choose \(j\ge0\) and identify two families of solutions
that satisfy the additional symmetry \(S_i(\phi)=S_i(\pi-\phi)\):
\begin{align}
  S_i\propto
  \left\lbrace
  \begin{array}{ll}
    \cos(j\phi)& j\textrm{ even or zero}\\
    \sin(j\phi)& j\textrm{ odd}\,.
  \end{array}
  \right.
  \label{phidependence}
\end{align}

The equation determining the \(\xi\)-dependent eigenfunction is
\begin{align}
\left(\xi T_{i}'\right)'-\left[\frac{j^2}{4\xi}+\Lambda_i\left(1-\xi\right)\right]T_{i}
&=0\,.
\label{xieq}
\end{align}
Following \citet[][appendix~A]{1989A&A...225..559K} (see also \citet[][section 7.31]{1927ode..book.....I}),
one looks for a solution of the form
\begin{align}
T_{i}&=\textrm{e}^{\lambda \xi}\xi^\alpha\sum_{n=0}^{\infty} c_{n} \xi^n\,.
\label{powerexpansion}
\end{align}
Inserting this into eq~(\ref{xieq}), shows that the choice
\(\lambda=-\sqrt{-\Lambda_i}\) is convenient, since it removes
the highest power of \(\xi\) in the term proportional to \(T_i\), leading to
a two-term recurrence relation for the \(c_n\). The
indicial equation is \(\alpha^2=j^2/4\), so that
there are two possible solutions 
\begin{equation}
\begin{aligned}
T^+_{i}&=\textrm{e}^{-\sqrt{-\Lambda}_i\xi}
  \sum_{n=0}^{\infty} \xi^n c^+_{n} & j\textrm{ even or zero}\\
  T^-_{i}&=\textrm{e}^{-\sqrt{-\Lambda}_i\xi}
  \sum_{n=0}^{\infty} \xi^{n+\frac{1}{2}}c^-_{n} & j\textrm{ odd.}
\end{aligned}
\label{powerexpansion2}
\end{equation}
Inserting (\ref{powerexpansion2}) into (\ref{xieq}) and equating
coefficients gives
\(j^2 c^+_0=0\) and \(\left(j^2-1\right)c^-_0=0\)
and, for \(n\ge0\), the recurrence relations
\begin{equation}
\begin{aligned}
  \label{recurrence}
c^+_{n+1}&=
\left( \frac{4\sqrt{-\Lambda_i}\left(2n+1-\sqrt{-\Lambda_i}\right)}{
  (2n+2)^2-j^2}\right)c^+_n  \\
c^-_{n+1}&=
\left(  \frac{4\sqrt{-\Lambda_i}\left(2n+2-\sqrt{-\Lambda_i}\right)}{
  (2n+3)^2-j^2}\right)c^-_{n} \,.
\end{aligned}
\end{equation}
Since
\(c^\pm_{n+1}/c^\pm_n\rightarrow 2\sqrt{-\Lambda_i}/n\), as \(n\rightarrow\infty\)
it follows that \(T^\pm_{i}\rightarrow\textrm{e}^{\sqrt{-\Lambda_i}\xi}\) as \(\xi\rightarrow\infty\), unless
the relevant series truncates at finite \(n\). Therefore, the eigenvalues satisfying
the boundary conditions of boundedness at \(\xi=0,\infty\)
are found  by
requiring truncation for, say, \(n>k\ge0\). Then,
\begin{align}
\Lambda_i&=\left\lbrace
\begin{array}{ll}
  -(2k+1)^2& j\textrm{ even or zero}, 0\le j\le 2k\\
  &\\
  -4(k+1)^2& j\textrm{ odd}, 1\le j\le 2k+1\,,
  \end{array}\right.
  \end{align}
  where \(k\) is a positive integer or zero.
  Thus, each index \(i\) corresponds to
  a pair of indices \(k\) and \(j\). The
  eigenvalues \(\Lambda_i\) depend on \(k\) and only
  the parity of \(j\). The corresponding eigenfunctions can be evaluated
  using the recurrence relations (\ref{recurrence}). The first few are
  listed in Table~\ref{table1}.
  \begin{table}
    \caption{2D-eigenfunctions:
      \(i\) uniquely identifies the
      eigenfunction \(Q_i\), \(j\) defines its \(\phi\)-dependence
      according to eq~(\ref{phidependence}), 
      \(k\) is the largest integer such that \(c_{k}\ne0\) in eq~(\ref{powerexpansion2}), 
      \(\Lambda_i\) is the eigenvalue determining the \(x\)-dependence
      associated with this
      eigenfunction in 
      eq~(\ref{expansion}). The (arbitrary) normalisation of the \(Q_i\) is
      \(c_{n}=1\), where \(n={j/2}\) (\(j\) even or zero),
  or \(n={(j-1)/2}\) (\(j\) odd) is the smallest integer for which
  \(c_n\ne0\).}
    \label{table1}
    \begin{tabular}{|l|l|l|l|l|}
    \hline
\(i\)     &\(j\)  &\(k\)    &\(\Lambda_i\)      &\(Q_{i}\)\\
    \hline\hline
&&&&\\
\(1\)     &\(0\)  &\(0\)    &\(-1\)       &\(\textrm{e}^{-{\xi}}\)
    \\
    \hline
&&&&\\
\(2\)     &\(1\)  &\(0\)    &\(-4\)        &\(\textrm{e}^{-2\xi}\sqrt{\xi}\sin(\phi)\)
    \\
    \hline
&&&&\\
\(3\)     &\(0\)  &\(1\)    &\(-9\)       & \(\textrm{e}^{-3{\xi}}\left(1-6{\xi}\right)\)
\\
&&&&\\
\(4\)     &\(2\)  &\(1\)    &\(-9\)       &\( \textrm{e}^{-3{\xi}}{\xi}\cos(2\phi)\)
\\
\hline
&&&&\\
\(5\)     &\(1\)  &\(1\)    &\(-16\)      & \(\textrm{e}^{-4{\xi}}\sqrt{{\xi}}\left(1-4{\xi}\right)\sin\phi\)
\\
&&&&\\
\(6\)     &\(3\)  &\(1\)    &\(-16\)      &\(\textrm{e}^{-4{\xi}}{\xi}^{3/2}\sin(3\phi)\)
\\
\hline
  \end{tabular}
\end{table}
    \section{Mapping}
\label{appendixB}
\begin{figure}
\begin{center}
\includegraphics[width = 0.5\textwidth]{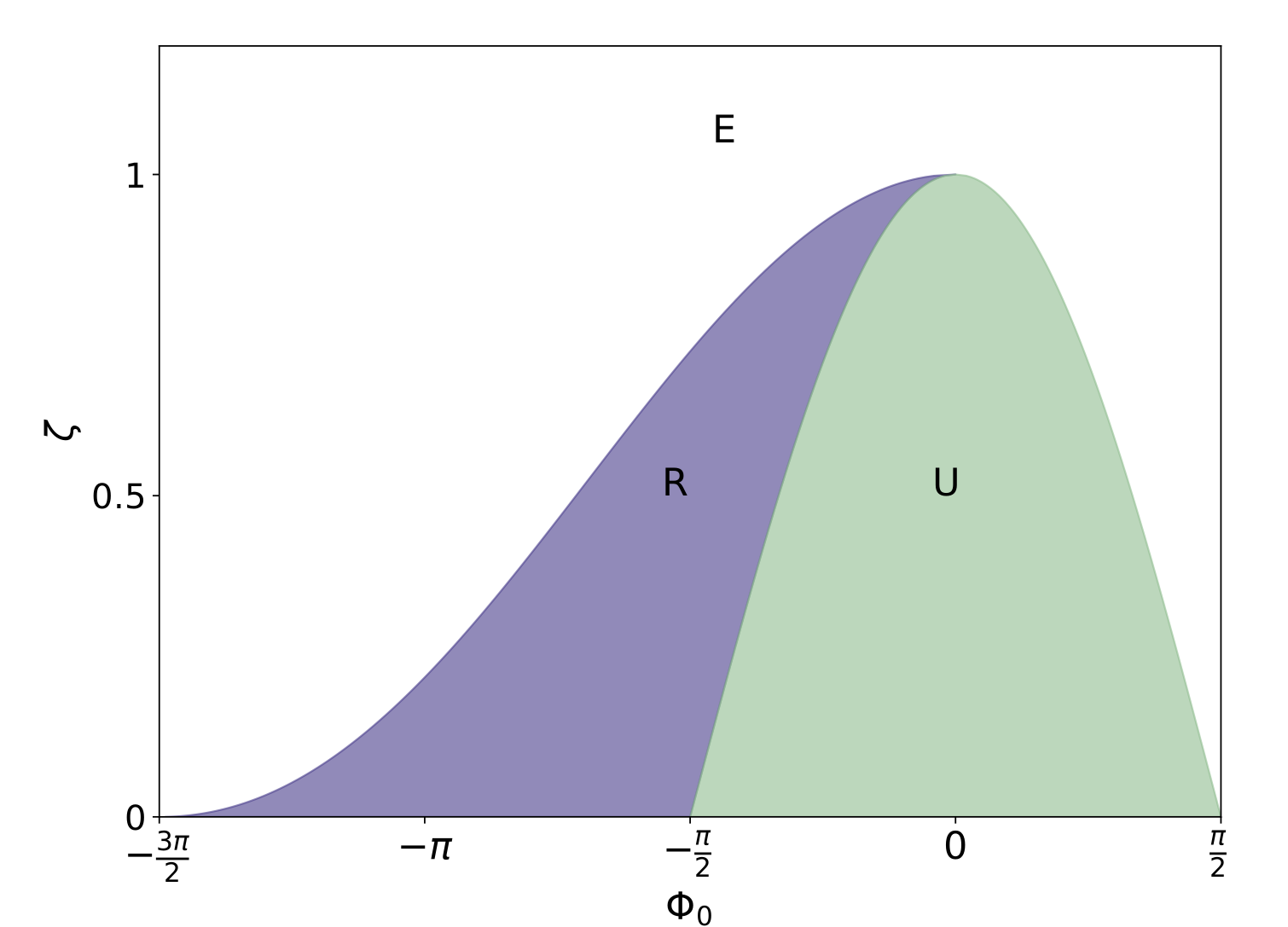}
\caption{The three sub-domains of particle phase space at the shock front in the \(\Phi_0 - \zeta\) plane: purple for \(\vec{p}\in R\), green for \(\vec{p}\in U\) and the remaining unshaded area for \(\vec{p}\in E\). Note that \(\zeta > \beta_{\rm d}\) .
	\label{fig:phasemaps}}
\end{center}
\end{figure}

For an ultrarelativistic particle (\(\gamma\gg\Gamma_{\rm s}\)),
membership of the three subdomains of momentum space
at the shock front --- upstream (\(U\)), returning (\(R\))
and escaping (\(E\)) --- 
of the particle phase space (\(A\)) is determined solely by
the direction of motion,
labelled by the stretched variable \(\xi\),
defined in (\ref{stretched}), and the azimuthal phase (with respect to
the shock normal) \(\phi\). Membership of \(U\) requires only
\(\xi<1\), but the requirements for \(R\) and \(E\) are more
complicated.  First, define the auxiliary parameter \(\zeta\)
to be the ratio of the shock speed \(c\beta_{\rm d}\) to the
component \(c\beta_{\bot}\) of the particle speed that is
perpendicular to the magnetic field, both seen from the frame in which
the downstream plasma is at rest. The computation of \(\zeta\)
from \((\xi,\phi)\)
involves a Lorentz boost from the rest frame of the upstream plasma to that
of the downstream plasma, followed by a change of axis of the polar coordinates
from the shock normal to the magnetic field. To lowest order in
\(1/\Gamma_{\rm s}\) one finds:
\begin{align}
  \zeta&=
         \frac{\beta_{\rm d}\left[\left(1+\beta_{\rm d}\right)+\left(1-\beta_{\rm d}\right)\xi\right]}
       {\sqrt{\left[\left(1+\beta_{\rm d}\right)+\left(1-\beta_{\rm d}\right)\xi\right]^2
         -4\xi\left(1-\beta_{\rm d}^2\right)\cos^2\phi}}
         \,\ge\,\beta_{\rm d}\,.
\end{align}
This quantity remains constant during the particle's residence
downstream. Clearly, \(\zeta>1\) implies that the particle cannot
recross the shock front (\(\vec{p}\in E\)) and the mapping
\({\cal M}\) is unity:
\begin{align}
  {\cal M}\vec{p}=\vec{p_+},\ \vec{p}\in E&:
                            \nonumber\\
                          &\gamma_+=\gamma\qquad\xi_+=\xi\qquad\phi_+=\phi\,.
\label{unity}
\end{align}
However, although \(\zeta>1\) is sufficient for
\(\vec{p}\in E\), it is not a necessary condition.
The distance \(d\), in units of the particle's
gyroradius, between the shock and a point on the trajectory depends on
\(\zeta\), the particle's phase \(\Phi\) (measured about the
downstream magnetic field, such that it is an increasing function of time
for a positively charged particle)
and the phase \(\Phi_0\) at which the
trajectory intersects the shock front:
\begin{align}
  d\left(\zeta,\Phi,\Phi_0\right)
  &= \zeta\left(\Phi-\Phi_0\right) -
    \left(\sin\Phi-\sin\Phi_0\right)\,.
\end{align}
From the Lorentz boost and coordinate transformation, one finds,
to lowest order in \(1/\Gamma_{\rm s}\),    
\begin{align}
  \Phi_0&=\textrm{atan2}\left[
          2\sin\phi\sqrt{\left(1-\beta_{\rm d}^2\right)\xi},\,
          \left(1+\beta_{\rm d}\right)-\left(1-\beta_{\rm d}\right)\xi\right]
+2n\pi          \,,
\end{align}
where, to simplify the discussion,
the integer \(n\) is to be chosen such that
\(-2\pi+\textrm{arccos}\zeta\le\Phi_0<\textrm{arccos}\zeta\).
For \(\vec{p}\notin U\), the distance \(d\) grows initially,
implying \(-2\pi+\textrm{arccos}\zeta<\Phi_0<-\textrm{arccos}\zeta\), and 
subsequently goes through an alternating series of maxima and
minima as the trajectory gyrates about the magnetic field.  The first
maximum is reached when \(\Phi=-\textrm{arccos}\zeta\) and the
subsequent minimum when \(\Phi=+\textrm{arccos}\zeta\). Therefore,
if \(\zeta<1\), a sufficient condition for the particle to escape
is that \(d>0\) at this minimum, i.~e.,
\begin{align}
d\left(\zeta,\textrm{arccos}\zeta,\Phi_0\right)&>0\,,
\end{align}
in which case \(\cal M\) is again the unit mapping (\ref{unity}).
If, on the other hand,
\(d<0\) at this point, then
\(\vec{p}\in R\), and the phase \(\Phi_{+}\),
with which the trajectory returns to the shock front, is given by the
single solution of the equation
\begin{align}
d\left(\zeta,\Phi_{+},\Phi_0\right)&=0\,,
\end{align}
with
\(-\textrm{arccos}\zeta<\Phi_{+}<\textrm{arccos}\zeta\).
Similarly, for \(\vec{p}\in U\), 
which implies \(-\textrm{arccos}\zeta<\Phi_0<\textrm{arccos}\zeta\), 
the phase \(\Phi_{-}\),
at which the trajectory
previously entered the downstream region, is the single solution of the
equation
\begin{align}
d\left(\zeta,\Phi_0,\Phi_{-}\right)&=0\,,
\end{align}
with
\(-2\pi+\textrm{arccos}\zeta<\Phi_{-}<-\textrm{arccos}\zeta\).
Thus, the boundaries between the different subdomains for particles in \(R\), \(E\) and \(U\) are uniquely defined
in the \(\Phi_0 - \zeta\) plane, as summarised in Figure \ref{fig:phasemaps}.

To find \({\cal M}\), it remains to transform the coordinates and
apply a Lorentz boost back to the
upstream frame, which gives, to lowest order in \(1/\Gamma_{\rm s}\),
\begin{align}
  {\cal M}\vec{p}=\vec{p_+},\ \vec{p}\in R&\textrm{ and }
  {\cal M}^{-1}\vec{p}=\vec{p_-},\ \vec{p}\in U:
                            \nonumber\\
                                          &\gamma_{\pm}=\gamma
\left(\zeta+\beta_{\rm d}\cos\Phi_{\pm}\right)                                            \frac{\left(1-\beta_{\rm d}\right)\xi+\left(1+\beta_{\rm d}\right)}
                                            {2\left(1-\beta_{\rm d}\right)\zeta}
  \nonumber\\
  &\xi_{\pm}=\frac{\left(1+\beta_{\rm d}\right)\left(\zeta-\beta_{\rm d}\cos\Phi_{\pm}\right)}
         {\left(1-\beta_{\rm d}\right)\left(\zeta+\beta_{\rm d}\cos\Phi_{\pm}\right)}
  \nonumber\\
          &\phi_{\pm}=\textrm{atan2}\left(\beta_{\rm d}\sin\Phi_{\pm},
            \sqrt{\zeta^2-\beta_{\rm d}^2}\right)\,.
\end{align}

\end{document}